\RequirePackage[2020-02-02]{latexrelease}
\documentclass[aps,twocolumn,10pt,aps,amsmath,amssymb,nofootinbib,superscriptaddress]{revtex4}
\usepackage{epsfig} 
\usepackage{amsmath} 
\usepackage{graphicx}
\usepackage{color}
\usepackage{float}
\usepackage{soul,xcolor}
\usepackage[font=scriptsize]{caption}
\usepackage{braket}
\usepackage{bbold}
\usepackage{subfig}
\usepackage{braket}
\usepackage{titlesec}
\usepackage{placeins}
\usepackage{hyperref}
\usepackage{caption}
\begin{document}
	\title{$CP$ violation due to a Majorana phase in two flavor neutrino oscillations with decays}

	\author{Khushboo Dixit}
	\email{kdixit@iitb.ac.in}
	\affiliation{Department of Physics, Indian Institute of Technology Bombay, Mumbai 400076, India}
	
	\author{Akhila Kumar Pradhan}
	\email{akhilpradhan@iitb.ac.in}
	\affiliation{Department of Physics, Indian Institute of Technology Bombay, Mumbai 400076, India}
	
	\author{S. Uma Sankar}
	\email{uma@phy.iitb.ac.in}
	\affiliation{Department of Physics, Indian Institute of Technology Bombay, Mumbai 400076, India}
	
	\date{\today}

\begin{abstract}
We study the conditions under which the Majorana phase of the two flavor neutrino mixing matrix appears in the oscillation probabilities and causes $CP$ violation. We find that the Majorana phase remains in the neutrino evolution equation if the neutrino decay eigenstates are not aligned with the mass eigenstates. We show that, in general, two kinds of $CP$ violation are possible: one due to the Majorana phase and the other due to the phase of the off-diagonal element of the neutrino decay matrix. We find that the $CP$ violating terms in the oscillation probabilities are also sensitive to neutrino mass ordering.
		
\end{abstract}
	\maketitle

\section{Introduction} \label{s:intro}

Neutrinos are the most intriguing particles in nature. They are the only known elementary neutral fermions. Even their fundamental nature, whether they are Dirac or Majorana fermions, is an open question. In the Standard Model (SM), neutrinos are massless. However, the discovery of neutrino oscillations showed that different neutrino flavors mix to form mass eigenstates and these states have tiny, nondegenerate masses, which are more than a million times smaller than the electron mass. The mechanism that gives rise to such tiny masses is also an open problem in particle physics. 

Since neutrinos are neutral, it is possible for them to be their own antiparticle, {\it i.e.,} they can be Majorana fermions. The mass term for Majorana neutrinos has a very different form compared to that of Dirac neutrinos and can be naturally made small via seesaw mechanism. Majorana masses violate lepton number by two units and lead to interesting signals such as neutrinoless double beta decay or same sign lepton pairs at colliders. If the neutrinos have Majorana masses, the mixing matrix connecting the flavor eigenstates to mass eigensates has extra phases (called Majorana phases). 

Neutrino oscillation probabilities depend on the elements of the mixing matrix and the mass-squared differences, in general. It is well established that vacuum oscillation probabilities do not depend on the Majorana phases \cite{Bilenky:1980cx,Schechter:1980gr,Doi:1980yb,Giunti:2010ec}. Matter effects arise due to neutrino propagation in dense matter and they modify the neutrino evolution and hence, neutrino oscillation probabilities. The matter effects can be due to pure SM interactions or they can include nonstandard interactions also. In both cases, it can be shown that the Majorana phases do not appear in the oscillation probabilities.

There are, however, some neutrino evolution equations for which the Majorana phases appear in neutrino oscillation probabilities. A new form of neutrino decoherence, with an off-diagonal term in the decoherence matrix was considered in Ref. \cite{Benatti:2001fa}. It was shown that the neutrino oscillation probabilities depend on Majorana phases in such a case. It was also shown that these probabilities are $CP$ violating \cite{Capolupo:2018hrp}. This leads us to the question of what other possibilities are there under which the Majorana phases appear in neutrino oscillation probabilities and lead to $CP$ violation.

	In this paper, we address the above question for the case of two flavor oscillations. Extension of the discussion to three flavor oscillations is straightforward. We consider the most general neutrino evolution Hamiltonian including decay terms. Then we identify the terms in this Hamiltonian which lead to the appearance of the Majorana phase in oscillation probabilities and discuss their $CP$ and $CPT$ properties.

\section{Vacuum neutrino oscillations}\label{dynamics}
 In this section, we briefly discuss the dynamics of two flavor neutrino oscillations in vacuum. In general, neutrino mass eigenstates $\nu_i$ mix via a unitary matrix to introduce flavor states $\nu_\alpha$ of neutrinos  
\begin{equation*}
	\mathbf{\nu}_{\alpha} = U~\mathbf{\nu}_i = O~U_{ph}~\mathbf{\nu}_i,
\end{equation*}
where $\nu_{\alpha} = (\nu_e~~~\nu_{\mu})^T$ and $\nu_i = (\nu_1~~~\nu_2)^T$, and  
\begin{equation}\label{mixing}
	O = \begin{pmatrix}
		\cos\theta   &\sin\theta\\
		-\sin\theta  &\cos\theta
	\end{pmatrix} ~~~~~~ U_{ph} = \begin{pmatrix}
		1      &0\\
		0  & e^{i\phi}
	\end{pmatrix}. 
\end{equation} 
The mixing matrix $U$ is parameterized in terms of the mixing angle $\theta$ and the Majorana phase $\phi$. We assume that two other phases are pulled out on the left and are absorbed in the flavor states. In the case of Dirac neutrinos, the phase $\phi$ gets absorbed in the neutrino mass eigenstates through rephasing and we are left with the orthogonal mixing matrix. Such rephasing cannot be done for Majorana neutrinos. 

Let us now consider the traditional diagonal Hamiltonian in mass basis that governs the time evolution of neutrino mass eigenstates,
\begin{align*}%
	\mathcal{H} &= \begin{pmatrix}
		a_1    &0\\
		0     &a_2
	\end{pmatrix}\\
 &= \frac{(a_1 + a_2)}{2}\begin{pmatrix}
	1    &0\\
	0     &1
\end{pmatrix} + \frac{(a_2-a_1)}{2} \begin{pmatrix}
		-1    & 0\\
		     0     & 1
	\end{pmatrix},
\end{align*}
where $a_1 = m_1^2/2E$ and $a_2 = m_2^2/2E$ where $m_i$ are the mass eigenvalues and $E$ is the energy of the neutrinos.  
Evolution equations in mass eigenbasis are
\begin{equation}\label{Ham}
	i\frac{d}{dt} \nu_i(t) = \left[\frac{(a_1+a_2)}{2} \sigma_0 - \frac{(a_2-a_1)}{2}\sigma_z\right]\nu_i(t),
\end{equation} 
where $\sigma_0$ is the $2\times2$ identity matrix and $\sigma_z$ is the diagonal Pauli matrix. In the flavor basis, Eq.~(\ref{Ham}) has the form 
\begin{equation}\label{Vacflv}
	i\frac{d}{dt} \nu_{\alpha}(t) = \left[\frac{(a_1+a_2)}{2} \sigma_0 - \frac{(a_2-a_1)}{2}O U_{ph}\sigma_zU_{ph}^\dagger O^T\right]\nu_{\alpha}(t).
\end{equation}
The term proportional to $\sigma_0$ makes no distinction between the flavors and hence is absent from the probabilities. Since, $U_{ph}$ and $\sigma_z$ are diagonal matrices, they commute; the $\sigma_z$ term simplifies to $ O\sigma_z O^T$ and the phase $\phi$ disappears from the evolution equation. The vacuum neutrino flavor transition probability is obtained as 
\begin{equation}\label{vac}
	P({\nu_e \rightarrow \nu_\mu}) = \sin^22\theta \sin^2\bigg(\frac{(a_2-a_1)t}{2}\bigg)\equiv P_{e\mu}^{\rm vac}. 
\end{equation}
The relations between different probabilities, $P_{ee}^{\rm vac} = 1-P_{e\mu}^{\rm vac} = P_{\mu\mu}^{\rm vac}$ and $P_{\mu e}^{\rm vac} = P_{e\mu}^{\rm vac}$, follow trivially. Expressing $a_i$ as $m_i^2/2E$ we get the standard expression for transition probability,
\begin{equation}\label{vac}
	P_{e\mu}^{\rm vac} = \sin^22\theta \sin^2\bigg(\frac{\Delta m^2 L}{4E}\bigg),
\end{equation}
	 where $L$ is the distance traveled by neutrinos.

	\section{Oscillations with general decay Hamiltonian} 
	 We now consider a neutrino Hamiltonian including decay terms. Such a Hamiltonian has the form 
\begin{equation}\label{nonHerm}
	 	\mathcal{H} = M - i \Gamma/2,
	 \end{equation}
	 where $M$ is the mass matrix and $\Gamma$ is the decay matrix. For a general two particle system these matrices are diagonal, {\it i.e.,} the mass eigenstates are also the decay eigenstates. However, for a system of two particles that can oscillate into each other, these matrices can have off-diagonal terms, as in the case of neutral meson system \cite{PKKabir,BigiSanda,Branco:1999fs}. In this work, we choose the matrices $M$ and $\Gamma/2$ to be of the form~\cite{Berryman:2014yoa,Chattopadhyay:2021eba}
\begin{equation}\label{GenHam1}
	  	\mathcal{M} = 
	  	\begin{pmatrix}
	  		a_1                &0\\
	  		0                & a_2
	  	\end{pmatrix},~~~ 
  	\Gamma/2 = 
  	\begin{pmatrix}
  		b_1                 &\frac{1}{2}\eta e^{i \xi}\\
  		\frac{1}{2}\eta e^{-i \xi}                & b_2
  	\end{pmatrix}.
	  \end{equation}
	\vspace{0.2cm}
 
  In Eq. (\ref{GenHam1}), the parameters $a_1, a_2, b_1, b_2, \eta$ and $\xi$ are real, with $a_2 - a_1 = \Delta m^2/2E$ depicting the frequency of neutrino oscillations and the rest causing decay of neutrino mass eigenstates. 
    The matrix $\Gamma$ needs to be positive semidefinite, {\it i.e.,} non-negative, which leads to the following constraints: $b_1, b_2 \geq 0$, and $\eta^2 \leq 4 b_1 b_2$. A beam of oscillating neutrinos has a spread of energies. For oscillating neutrinos, this spread is large enough such that both $\nu_1$ and $\nu_2$ are on mass shell. Because of this spread, both $\nu_1$ and $\nu_2$ can decay into the same set of final states with the same energies. This possibility gives rise to the off-diagonal term $\Gamma_{12}$.
    
	 Neutrino evolution through this Hamiltonian describes both  neutrino oscillation as well as neutrino decay. If $\Gamma$ is diagonal ($\eta = 0$), {\it i.e.,} the decay eigenbasis is the same as the mass eigenbasis. In such a case it is straightforward to show that the Majorana phase $\phi$ disappears from neutrino evolution equations through a discussion similar to that of the previous section. We now consider the case where $\Gamma$ is non-diagonal ($\eta \neq 0$), {\it i.e.,} the mass eigenstates are {\it not} decay eigenstates. 
	  In this case, the evolution equation in the mass eigenbasis takes the form
	   \begin{eqnarray}\label{Vacmassdec}
	  	i\frac{d}{dt} \nu_{i}(t) &=&  \left[\frac{(a_1+a_2)}{2}\sigma_0 -\frac{(a_2-a_1)}{2}\sigma_z\right. \nonumber \\
	  	&&-\left. \frac{i}{2}\left((b_1 + b_2)\sigma_0+ \vec{\sigma}.\vec{\Gamma}\right)\right]~\nu_{i}(t),
	  \end{eqnarray}
  where $\vec{\Gamma} = [\eta \cos \xi, -\eta \sin \xi, -(b_2-b_1)]$. 
	  Transforming this equation to flavor basis we get
	   \begin{eqnarray}\label{Vacflvdec}
	  	i\frac{d}{dt} \nu_{\alpha}(t) &=&  \left[\frac{(a_1+a_2)}{2}\sigma_0 -\frac{(a_2-a_1)}{2}O \sigma_z O^T\right. \nonumber \\
	  	&&-\left. \frac{i}{2} (b_1 + b_2)\sigma_0 - \frac{i}{2} O U_{ph} (\vec{\sigma}.\vec{\Gamma}) U_{ph}^{\dagger}O^T\right]\nu_{\alpha}(t).\nonumber\\ 
	  \end{eqnarray}
  Since $\sigma_x$ and $\sigma_y$ do not commute with $U_{ph}$ matrix, the phase $\phi$ remains in the evolution equation.

  The time evolution operator for neutrinos in the mass eigenbasis is $\mathcal{U} = e^{-i \mathcal{H} t}$. This matrix can be expanded in the basis spanned by $\sigma_0$ and Pauli matrices \cite{Nielsen}. This expansion is parameterized by a complex four-vector $n_\mu \equiv (n_0,\vec{n})$, whose components are given by $n_\mu = Tr[(- i \mathcal{H} t).\sigma_\mu]/2$. Explicitly, they are expressed in terms of the parameters of $\mathcal{H}$ as
  \begin{eqnarray}
  	n_0 = -\frac{i}{2}(a_1 + a_2)t - \frac{1}{2}(b_1 + b_2)t, \nonumber\\
  	n_x = -\frac{1}{2} (\eta \cos \xi)t, \nonumber\\
  	n_y = \frac{1}{2} (\eta \sin \xi)t, \nonumber\\
  	n_z = \frac{i}{2}(a_2 - a_1)t + \frac{1}{2}(b_2 - b_1)t.
  \end{eqnarray} 
In terms of these components the evolution matrix $\mathcal{U}$ is
   \begin{equation}\label{Um}
   	\mathcal{U} =  e^{n_0} \bigg[\cosh n ~\sigma_0 + \frac{\vec{n}.\vec{\sigma}}{n} \sinh n \bigg],
   \end{equation}
   where 
   \begin{equation}
   	n = \sqrt{n_x^2 + n_y^2 + n_z^2} = \frac{t}{2}\sqrt{\eta^2 - (a_2 - a_1 - i (b_2 - b_1))^2}.
   \end{equation} 
 
 The evolution matrix in flavor basis can be obtained through the transformation $\mathcal{U}_f = U \mathcal{U} U^{-1}$, where $U$ is defined in Eq. (\ref{mixing}). 
 Oscillation probabilities can be obtained as
   \begin{equation*}
   	P_{\alpha\beta} = |(\mathcal{U}_f)_{\alpha\beta}|^2.
   \end{equation*}
    
   The general probability expressions with all the decay parameters nonzero are quite complicated. In this article, we are interested in how the probabilities depend on the Majorana phase $\phi$. To illustrate this, we consider the probability expressions in the limit $b_1 = b = b_2$ and $\eta \ll |a_2-a_1|$. For convenience we define 
   \small
   \begin{eqnarray}
   	\mathcal{A} &=& \frac{\sin (2 \theta ) \sin \left[\left(a_2-a_1\right)t\right]}{(a_2-a_1)},~~ \nonumber \\
   	 \mathcal{B} &=& \frac{\sin (2 \theta ) \sin ^2\left[\frac{1}{2} t (a_2-a_1)\right]}{(a_2-a_1)}.
   \end{eqnarray}
   
\normalsize
    Neglecting terms of $\mathcal{O}(\eta^2)$ and higher order, we get the survival probabilities as 
    \begin{eqnarray}\label{nuprob1}
    	P_{ee} = e^{-2bt} \left(P_{ee}^{\rm vac} - \eta \cos(\xi-\phi) \mathcal{A}\right)~~ {\rm and}\nonumber \\
    	P_{\mu\mu} =  e^{-2bt} \left(P_{\mu\mu}^{\rm vac} + \eta \cos(\xi-\phi) \mathcal{A}\right)~~~~~~
    \end{eqnarray}
and the oscillation probabilities as
\begin{eqnarray}\label{nuprob2}
	P_{e\mu} = e^{-2bt} \left(P_{e\mu}^{\rm vac} + 2\eta \sin(\xi-\phi) \mathcal{B}\right)~~ {\rm and}\nonumber \\ 
	P_{\mu e} =  e^{-2bt} \left(P_{\mu e}^{\rm vac} - 2\eta \sin(\xi-\phi) \mathcal{B}\right).~~~~~~
\end{eqnarray}
Hence, we see that the Majorana phase $\phi$ appears in the probability expressions if the neutrino evolution equation contains the off-diagonal term of the decay matrix $\Gamma_{12} \propto \eta$. The presence of this term also violates the equalities $P_{\mu\mu} = P_{ee}$ and $P_{\mu e} = P_{e\mu}$ that we see in the case of two flavor vacuum oscillations. In addition, we note that the terms with $\mathcal{B}$, present in oscillation probabilities, have opposite signs for the two cases $a_2>a_1$ ($m_2>m_1$) and $a_2<a_1$ ($m_2<m_1$); that is, the oscillation probability is sensitive to the mass hierarchy.

We now consider the oscillations of antineutrinos. We assume $CPT$ conservation which implies the following relations for the mass and decay matrices \cite{Berryman:2014yoa}
\begin{equation}
	\bar{M} = M ~~{\rm and}~~ \bar{\Gamma} = \Gamma^\ast.
\end{equation}
Hence, antineutrino probabilities expressions can be obtained by making the substitutions $\phi \rightarrow -\phi$ and $\xi \rightarrow -\xi$ in the neutrino probability expressions. Explicitly these expressions are 
\begin{eqnarray}\label{antinuprob1}
	P_{\bar{e}\bar{e}} = e^{-2bt} \left(P_{\bar{e}\bar{e}}^{\rm vac} - \eta \cos(\xi-\phi) \mathcal{A}\right)\nonumber \\ 
	P_{\bar{\mu}\bar{\mu}} =  e^{-2bt} \left(P_{\bar{\mu}\bar{\mu}}^{\rm vac} + \eta \cos(\xi-\phi) \mathcal{A}\right)
\end{eqnarray}
and 
\begin{eqnarray}\label{antinuprob2}
	P_{\bar{e}\bar{\mu}} = e^{-2bt} \left(P_{\bar{e}\bar{\mu}}^{\rm vac} - 2\eta \sin(\xi-\phi) \mathcal{B}\right)\nonumber \\ 
	P_{\bar{\mu}\bar{e}} =  e^{-2bt} \left(P_{\bar{\mu}\bar{e}}^{\rm vac} + 2\eta \sin(\xi-\phi) \mathcal{B}\right).
\end{eqnarray}
Since we assumed $CPT$ invariance we find $P_{\bar{e}\bar{e}} = P_{ee}$, $P_{\bar{\mu}\bar{\mu}} = P_{\mu\mu}$ and $P_{\bar{\mu}\bar{e}} = P_{e\mu}$. However, there is $CP$ violation ($P_{\bar{e}\bar{\mu}} \neq P_{e\mu}$) and $T$ violation ($P_{\mu e} \neq P_{e\mu}$).

    \section{Results and Discussions}\label{discussions}
	In this section, we discuss our results. In transforming the evolution equation from the mass eigenbasis to the flavor eigenbasis, we get the matrix product $U_{ph} \mathcal{H} U_{ph}^\dagger$, where $\mathcal{H}$ is the Hamiltonian in mass eigenbasis. The diagonal phase matrix $U_{ph}$ commutes with $\mathcal{H}$ whenever $\mathcal{H}$ is diagonal. This is true for the usual neutrino oscillations and for the case where the mass eigenstates are also the decay eigenstates. In such situations, $U_{ph}$ matrix drops out from the neutrino evolution equation, which in turn leads to the oscillation probabilities being independent of the Majorana phase $\phi$. However, when the decay eigenstates are not aligned with the mass eigenstates, there is an off-diagonal term in the decay matrix $\Gamma$ of $\mathcal{H}$. Since $\mathcal{H}$ is no longer diagonal, it does not commute with $U_{ph}$, thus leading to the presence of $\phi$ in the evolution equation and in the probabilities. In addition to the off-diagonal dissipator discussed in \cite{Benatti:2001fa}, the off-diagonal decay matrix is another possible source for the appearance of Majorana phases in the oscillation probabilities and the corresponding $CP$ violation. 
	
	The off-diagonal term of the decay matrix $\Gamma$ has the general form $\eta e^{i \xi}$ and the $CP$ violating term in the oscillation probabilities is proportional to $\eta \sin(\xi - \phi)$. We distinguish different forms of $CP$ violation based on the values of the phases $\xi$ and $\phi$.
	\begin{itemize}
	    \item For $\eta \neq 0$ and  $\xi = 0$ the decay matrix $\Gamma$ is real and is $CP$ conserving. In such a case, we need $\phi \neq 0$ for $CP$ violation. We call this $CP$ violation in mass because $\phi$ arises due to the diagonalization of the complex mass matrix.
	    \item If $\phi = 0$, it is still possible to have $CP$ violation if $\eta \neq 0$ and $\xi \neq 0$. We call this $CP$ violation in decay because the $CP$ violating phase $\xi$ comes from the decay matrix.
	    \item The most general possibility is $\eta \neq 0$, $\xi \neq 0$ and $\phi \neq 0$. In this case, we have $CP$ violation due to both mass and decay provided $\phi \neq \xi$. 
	\end{itemize}
        We see that $\eta \neq 0$ in all the above three cases. However, a nonzero value of $\eta$ is a necessary condition for $CP$ violation but not a sufficient condition. 
	For the two special cases, (a) $\phi = 0 = \xi$ and (b) $\phi = \xi$, there is no $CP$ violation even when $\eta \neq 0$. In these two cases, the $CP$ violating terms vanish and the flavor conversion probabilities are the same as the vacuum probabilities multiplied by the decay term. However, the presence of a nonzero value of $\eta$ is discernible in the survival probabilities.

  We now briefly discuss the values of parameters for which the effects described in this work are likely to be observable. A bound of $\tau_{\nu}\geq 5.7 \times 10^{5}$ s ($m_{\nu}/{\rm eV}$) is derived from the neutrino data of Supernova 1987A \cite{Frieman:1987as}, which leads to $\Gamma_{\nu} \equiv b \approx 10^{-21}$ eV for a neutrino of mass 1 eV. We take $\eta = b$, which satisfies the semipositivity constraint $\eta \leq 2 b$. The new effects considered in this work are of order $\eta/(a_2-a_1) = \eta E/\Delta m^2$. These effects are of order 10\% for $\Delta m^2 \approx 10^{-4} \,{\rm eV^2}$ if $E \approx 10^{16}$ eV or $10^{7}$ GeV. That is, ultrahigh energy neutrinos from astrophysical sources provide a platform to study the effect of the off-diagonal decay term considered here.

	 \section{Summary and Conclusions}\label{conclusions}
	 In this article, we point out scenarios in which the Majorana phase can appear in neutrino oscillation probabilities that also cause $CP$ violation. We did this analysis for two flavor oscillations, but the extension of this work for three flavor oscillations is straightforward. $CP$ violation in neutrino oscillations requires complex values of neutrino mixing matrix. In the case of standard two flavor oscillations, the phases of this matrix, including the Majorana phase, drop out of the evolution equation and there is no $CP$ violation. In this work, we have shown that the Majorana phase in two flavor mixing remains in the evolution equation and causes $CP$ violation provided the neutrinos decay and the decay eigenstates are not the same as mass eigenstates. This requires the decay matrix $\Gamma$ to have an off-diagonal term $\Gamma_{12}$. We have shown that two types of $CP$ violation are possible: (a) that due to the Majorana phase $\phi$, which we call $CP$ violation in mass, and (b) that due to the phase $\xi$ of $\Gamma_{12}$, which we call $CP$ violation in decay. The $CP$ violating term in the oscillation probability is also sensitive to the neutrino mass ordering. In the two special cases, when $\phi$ and $\xi$ are equal to each other or when both are zero, there is no $CP$ violation even if the decay eigenstates are different from the mass eigenstates. In such a situation, the flavor conversion probabilities are insensitive to 
	 off-diagonal elements of $\Gamma$ but the flavor survival probabilities do depend on them.

	\section*{Acknowledgments}
	 K.~D. and S.~U.~S. thank the Ministry of Education,
	 Government of India, for financial support through
	 Institute of Eminence funding.

\bibliographystyle{simple}

\end{document}